\documentclass{article}
\usepackage{spconf,amsmath,graphicx}
\usepackage{amsmath,amssymb,amsfonts}
\usepackage{algorithmic}
\usepackage{graphicx}
\usepackage{textcomp}
\usepackage{booktabs}
\usepackage{multirow}
\usepackage{color}
\usepackage{cite}
\usepackage{booktabs} 
\usepackage{supertabular}
\usepackage{xspace}
\usepackage{url}
\usepackage{soul}
\newcommand{\methodname}{FAN-Net\xspace}

\newcommand{\vct}[1]{\boldsymbol{#1}} 
\newcommand{\mat}[1]{\boldsymbol{#1}} 
\newcommand{\ie}{\emph{i.e.}\xspace} 
 
\newcommand{\etc}{\emph{etc}\xspace}

\title{FAN-Net: Fourier-based Adaptive Normalization for Cross-Domain Stroke Lesion Segmentation}
\name{Weiyi~Yu$^{1}$ \quad Yiming~Lei$^{2,\dagger}$ \quad Hongming~Shan$^{1,3}$ \thanks{$\dagger$: Corresponding Author.  This work was supported in part by Natural Science Foundation of Shanghai (No. 21ZR1403600), National Natural Science Foundation of China (No. 62101136), China Postdoctoral Science Foundation (No. 2022TQ0069) , Shanghai Municipal of Science and Technology Project (No. 20JC1419500), and  Shanghai Center for Brain Science and Brain-inspired Technology.}}
\address{
$^{1}$ Institute of Science and Technology for Brain-inspired Intelligence\\
$^2$ Shanghai Key Lab of Intelligent Information Processing, School of Computer Science\\
Fudan University, Shanghai 200433, China\\
$^3$ Shanghai Center for Brain Science and Brain-Inspired Technology, Shanghai 201210, China
}

\begin{document}
%
\maketitle
%

\begin{abstract}\label{sec:abs}
Since stroke is the main cause of various cerebrovascular diseases, deep learning-based stroke lesion segmentation on magnetic resonance (MR) images has attracted considerable attention. However, the existing methods often neglect the domain shift among MR images collected from different sites, which has limited performance improvement. To address this problem, we intend to change style information without affecting high-level semantics via adaptively changing the low-frequency amplitude components of the Fourier transform so as to enhance model robustness to varying domains. Thus, we propose a novel \methodname, a U-Net--based segmentation network incorporated with a Fourier-based adaptive normalization (FAN) and a domain classifier with a gradient reversal layer.
The FAN module is tailored for learning adaptive affine parameters for the amplitude components of different domains, which can dynamically normalize the style information of source images. Then, the domain classifier provides domain-agnostic knowledge to endow FAN with strong domain generalizability.
The experimental results on the ATLAS dataset, which consists of MR images from 9 sites, show the superior performance of the proposed \methodname compared with baseline methods.

\end{abstract}


\section{Introduction}\label{sec:intro}

Stroke is the main cause of death worldwide~\cite{2020Heart}. Traditionally, to examine the patients after a stroke, medical experts manually segment stroke lesions on T1-weighted magnetic resonance (MR) images, and this requires a costly workload. Therefore, it triggers a popular research topic over the past decade---automatic stroke lesion segmentation.

Convolutional neural networks (CNNs) are commonly used models in medical image segmentation~\cite{malhotra2022deep}. The current works related to stroke lesion segmentation can be roughly categorized into two categories, i.e., 2D CNNs and 3D CNNs.
The 2D CNN-based methods 
can generate a large number of independent samples for training but hardly capture the inter-slice relationships. Therefore, X-Net with a non-local attention module~\cite{2019X} and MSDF-Net with a multiscale fusion module~\cite{2019MSDF} were proposed to learn more diverse relationships among slices. By contrast, 3D CNNs are able to extract inter-slice and intra-slice relationships simultaneously; nevertheless, they require more 3D samples to prevent overfitting. Thus, D-UNet~\cite{2019D}, \etc.~\cite{2019A, DBLP:conf/miccai/ZhangWLCWT20, DBLP:journals/sncs/BasakHR21} combined the 2D and 3D CNNs to achieve a trade-off between them. In addition, efforts have also been made on the training strategy. 
For instance, the MI-UNet~\cite{2020MI} introduced brain parcellations as prior knowledge, which is useful but time-consuming.

\begin{figure}[t]
\centering
\includegraphics[width=0.7
\linewidth]{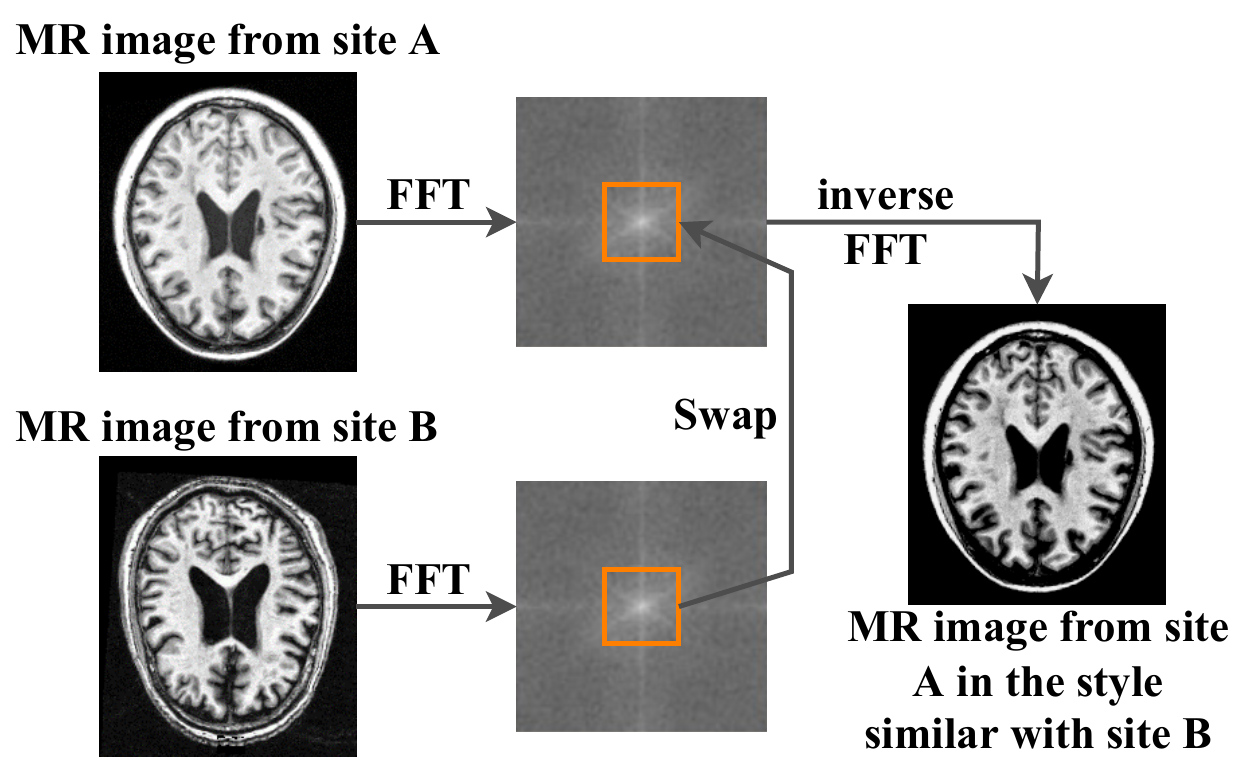}
\vspace{-4mm}
\caption{Example that proves the low-frequency amplitude component contains the style information.
}
\label{motivation}
\end{figure}

The methods mentioned above have neglected the domain diversity inherited in MR images from different sites. More specifically, the cross-domain differences involve MR scanners, imaging protocols, and patient populations, which greatly affect model generalization. Fortunately, domain generalization is an effective technique that helps machine learning models perform well on medical images of $\textit{unseen}$ domains~\cite{sun2021multi},
and gradient reversal layer (GRL)~\cite{ganin2015unsupervised} has been testified as an effective way of reducing cross-domain variance~\cite{yu}. Generally, GRL is combined with a domain classifier and tends to confuse this classifier by directly maximizing the classification loss using gradient reversing. 

\begin{figure*}[ht]
\centering
\includegraphics[width=0.85\linewidth]{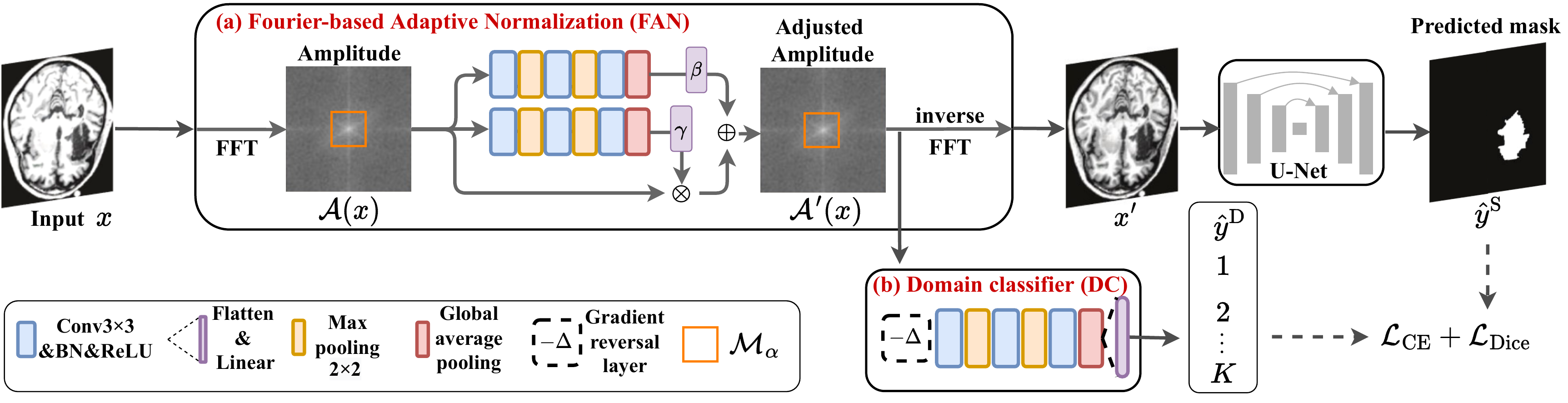}
\vspace{-5mm}
\caption{Illustration of the proposed \methodname for brain stroke lesion segmentation.
(\textbf{a}) Fourier-based adaptive normalization (FAN) standardizes the MR images into a domain-unrelated style. (\textbf{b}) Domain classifier with gradient reversal layer ($-\vct{\Delta}$).}
\vspace{-4mm}
\label{fig:architecture}
\end{figure*}

Hence, we need to model the domain knowledge or domain style information explicitly. It is well-known that the phase component of the Fourier spectrum preserves high-level semantic structures, and the amplitude component contains style information statistics~\cite{oppenheim1979phase, piotrowski1982demonstration}. Since one pixel in the image space simultaneously relates to the phase and amplitude components of the Fourier spectrum, the methods directly operating on the image space cannot change the global style information without affecting the semantic structures. However, processing frequency space can easily preserve the semantic features when changing style information. For example in Fig.~\ref{motivation}, an MR image from site A can be converted to a style that is similar to site B by substituting the low-frequency component of the amplitude spectrum. 
Thus, the style information can be represented by the low-frequency amplitude components ~\cite{yang2020fda, xu2021fourier, jiang2022harmofl}. 

Motivated by this perspective in the frequency space, we can change style information while preserving high-level semantic structures through a domain-agnostic amplitude.
Hence, we propose \methodname to tackle the problem of \textit{unseen} domain generalization in stroke lesion segmentation. Specifically, we propose a Fourier-based adaptive normalization (FAN) module to dynamically learn adaptive affine parameters for the amplitude components, which could convert the input MR images into a domain-unrelated style. Then, the style-transferred source image also acts as the input of a domain classifier, which further guarantees FAN to minimize cross-domain diversity through GRL. Finally, it is fed into U-Net for segmentation. The experiments conducted on the Anatomical Tracings of Lesions After Stroke (ATLAS) dataset~\cite{2018A} have illustrated the superior results obtained by \methodname, note that the FAN-Net outperforms baseline methods without large memory and computational overheads.

\section{Methodology}\label{sec:method}
Let $\mat{x} \in \mathbb{R}^{H \times W}$ be a source image with domain label $y^{\text{D}} \in \{1,2,\ldots, K\}$ where \emph{K} is the number of source classes, and the corresponding segmentation ground-truth is $\mat{y}^{\text{S}}\in \mathbb{R}^{H \times W}$. Our aim is to predict the segmentation result of an input from any target domain. The superscripts ``D'' and ``S'' denote domain and segmentation, respectively.

\subsection{Fourier-based Adaptive Normalization}

Since the amplitude component in Fourier Transform contains low-level statistics, swap, and linear decomposition are proposed to change the style information of MR images without affecting high-level semantics. However, the existing Fourier-based methods do not consider how to enable the model to be robust to an unseen domain. It motivates us to propose \methodname, a dynamic neural network 
that can learn how to normalize the input images to reduce cross-domain discrepancies. FAN leverages two affine transformation parameters learned from the amplitude components of MR images to convert them into a domain-agnostic style.

For a 2D image $x$, its Fourier transform $\mathcal{F}(x)$ is formulated as:
\begin{equation}
\mathcal{F}(x)(u, v)=\sum_{h=0}^{H-1} \sum_{w=0}^{W-1} x(h, w) e^{-j 2 \pi\left(\frac{h}{H} u+\frac{w}{W} v\right)},
\end{equation}
where $u$ and $v$ are the coordinates in the frequency space.
Then, let $\mathcal{F}^{-1}(x)$ denote the inverse Fourier transform, then $\mathcal{F}(x)$ and $\mathcal{F}^{-1}(x)$ can be calculated with the Fast Fourier Transform (FFT) algorithm~\cite{nussbaumer1981fast}, and the amplitude and phase components can be represented as:
\begin{align}
\mathcal{A}(x)(u, v)&=\left[R^2(x)(u, v)+I^2(x)(u, v)\right]^{1 / 2},\\
\mathcal{P}(x)(u, v)&=\arctan \left[\frac{I(x)(u, v)}{R(x)(u, v)}\right],
\end{align}
where $R(x)$ and $I(x)$ represent the real and imaginary parts of $\mathcal{F}(x)$, respectively.

Further, we denote $M_\alpha$ as a binary mask:
\begin{equation}
M_\alpha(h, w)\!=\!
\begin{cases}
1,  \text {if} (h, w) \!\in\![-\alpha H\!:\! \alpha H,-\alpha W\!:\! \alpha W] \\
0,  \text {otherwise}
\end{cases},
\end{equation}
where we treat the center point of the image as $(0,0)$, and $\alpha \in(0,1)$ is set manually.

We assume the affine transformation parameters $\gamma$ and $\beta$ are learned from the amplitude component of the MR image. The adjusted amplitude component can be represented as:
\begin{equation}
\mathcal{A'}(x) = \gamma \times {A}(x) + \beta, \label{eq:alpha}
\end{equation}
where $\beta = g_{\beta}(\mathcal{A}(x); \mat{\theta}_{\beta})$, and $\gamma = g_{\gamma}(\mathcal{A}(x); \mat{\theta}_{\gamma})$, $g_{\beta}$ and $g_{\gamma}$ are two lightweight networks for learning $\beta$ and $\gamma$, respectively. The structures of $g_{\beta}$ and $g_{\gamma}$ are the same but their parameters are independent. As illustrated in Fig.~\ref{fig:architecture}(a), $\beta$ and $\gamma$ are the outputs of global average pooling layer. Then the output of FAN can be formalized as:
\begin{equation}
x' =\mathcal{F}^{-1}\left(\left[M_\alpha \circ \mathcal{A'}(x)+\left(1-M_\alpha\right) \circ \mathcal{A}(x)\right], \mathcal{P}(x)\right),
\label{x'}
\end{equation}
where $x'$ represents the image dynamically converted from the original $x$, $\circ$ denotes element-wise multiplication. Note that the parameters of FAN, \ie, $\mat{\theta}_{\beta}$ and $\mat{\theta}_{\gamma}$, can be updated through gradient back-propagation.

\subsection{Domain-agnostic Knowledge Learning}
\label{detail}

We have designed the learning scheme for adaptive affine parameters $\beta$ and $\gamma$, the FAN is still sensitive to different domain knowledge. In order to endow $\beta$ and $\gamma$ with domain-unrelated knowledge, we further identify $\mathcal{A'}(x)$ using a domain classifier (DC) with a GRL. Here, we use the cross-entropy (CE) loss to train the domain classifier:
$
\mathcal{L}_{\text{CE}}=- \sum_{k=1}^{K} y^{\text{D}}_{k} \log \hat{y}^{\text{D}}_{k}
$, where $y^{\text{D}}_{k}$ represents the k-th domain class, and $\hat{y}^{\text{D}}_{k}$ denotes the predicted probability of $k$-th domain class. 

Due to the gradient reversing by GRL, the DC practically maximizes the CE loss so that the gradients $\frac{\partial{\mathcal{L}_{\text{CE}}}}{\partial{\mat{\theta}_{\beta}}}$ and $\frac{\partial{\mathcal{L}_{\text{CE}}}}{\partial{\mat{\theta}_{\gamma}}}$ are toward the same direction of gradients of DC, \ie, $\beta$ and $\gamma$ are learned to be insensitive to different domain knowledge. Hence, $x'$ in Eq.~(\ref{x'}) will have no specific domain information. Consequently, after training with multiple domains of source images, the FAN can convert style information of any domain to an unified form without affecting semantic structures.

\subsection{Loss Functions}
When we guarantee that the $x{'}$ is domain-agnostic, it can be used as an input of a U-Net~\cite{2015U} segmentation network. Dice loss is used for segmentation $\mathcal{L}_{\text{Dice}}(\hat{y}^{\text{S}}, y^{\text{S}})$, where $\hat{y}^{\text{S}}$ is the output of U-Net.
Then the total loss function is defined as:
\begin{equation}
\mathcal{L}= \mathcal{L}_{\text{Dice}} + \lambda \mathcal{L}_{\text{CE}},
\label{total_loss}
\end{equation}
where $\lambda$ decides the weight of $\mathcal{L}_{\text{CE}}$ and is empirically set as $1$. Minimizing $\mathcal{L}_{\text{Dice}}$ triggers domain-agnostic segmentation results, which is attributed to FAN and domain classifier.

\section{Experiments}\label{sec:exp}
\subsection{Dataset and Implementation Details}\label{sec:exp:dataset}
ATLAS~\cite{2018A} is the unique and high-quality open-source dataset of 229 patients' T1-weighted MR images (version: v1.2). For each subject, $189$ T1-weighted images were acquired and normalized into the standard space (MNI-152 template). The original size of images is $233 \times 197$, and all the images were cropped into $224 \times 192$. The size of lesions ranges from $10$ to $2.838\times10^5$ mm$^3$, and the ground-truth segmentation masks were manually segmented by specialists. 

What's more, this dataset was collected from nine sites, and more specific details are listed in the supplementary materials. Population variation leads to different vascular territories, meanwhile, the MR images were acquired by various types of 3T MR scanners and imaging protocols. These factors result in substantial challenges in cross-domain learning.

The \methodname is implemented with PyTorch and trained on an NVIDIA V100 Tensor Core GPU. In our experiments, we use SGD optimizer, the mini-batch size is 8, and a total of 50 epochs. The learning rate is set to 0.001 initially, with a 0.04 weight decay after each epoch. The parameter $\alpha$ of the binary mask $M_\alpha$ is set to 0.1. More details are provided in the supplementary material (\url{https://github.com/ymLeiFDU/Supplementary-for-FAN-Net/blob/main/Supplementary-for-FAN-Net.pdf}).

\subsection{Performance Comparisons}\label{sec:exp:comparison}

All the experiments were conducted under the ``leave-one-site-out'' setting, which regards one site as the test/target set and the other sites as the training/source set. We compare FAN-Net with baseline methods including U-Net~\cite{2015U}, ResUNet~\cite{2017Road}, PSPNet~\cite{DBLP:conf/cvpr/ZhaoSQWJ17}, DeepLabv3+~\cite{DBLP:conf/eccv/ChenZPSA18}, X-Net~\cite{2019X}, U-Net3+~\cite{DBLP:conf/icassp/HuangLTHZIHCW20}, nnU-Net~\cite{isensee2021nnu}, and Unlearning\cite{dinsdale2021deep}. Note that all the images were preprocessed by z-score normalization.

\begin{table*}[ht]\small
\renewcommand{\arraystretch}{1.0}
\caption{Quantitative comparisons under the leave-one-site-out setting. 
\#Par.: the number of model parameters; Mem.: total GPU memory of the model; MACC: multiply-accumulate operations; and FLOPs: floating-point operations.}
\label{tab:performance_comparison}
\centering
\begin{tabular*}{1.0\linewidth}{@{\extracolsep{\fill}}rcccrrrr}

\toprule
\textbf{Method} &\textbf{Dice} &\textbf{Recall} &\textbf{F1-score} & \#\textbf{Par.} [M] & \textbf{Mem.} [MB] & \textbf{MACC} [G] & \textbf{FLOPs} [G]\\

\midrule
U-Net~\cite{2015U} & 0.4712 $\pm$ 0.1952 & 0.4315 $\pm$ 0.1931 & 0.4864 $\pm$ 0.2161 & 28.94 & 260.20 & 63.21 & 31.63\\

ResUNet~\cite{2017Road} & 0.4780 $\pm$ 0.1952 & 0.4693 $\pm$ 0.1931 & 0.5322 $\pm$ 0.1846 & 28.94 & 260.20 & 63.20 & 31.63\\

PSPNet~\cite{DBLP:conf/cvpr/ZhaoSQWJ17} & 0.4318 $\pm$ 0.2054 & 0.3813 $\pm$ 0.1792 & 0.3921 $\pm$ 0.1948 & 38.28 & 261.91 & 65.07 & 32.56\\

Deeplabv3+~\cite{DBLP:conf/eccv/ChenZPSA18} & 0.4639 $\pm$ 0.2077 & 0.4594 $\pm$ 0.2181 & 0.4714 $\pm$ 0.1840 & 59.33 & 171.63 & 28.98 & 14.50\\

X-Net~\cite{2019X} & 0.5083 $\pm$ 0.1926 & 0.4954 $\pm$ 0.1844 & 0.5179 $\pm$ 0.1896 & \textbf{15.05} & 915.67 & 40.49 & 20.33\\

U-Net3+~\cite{DBLP:conf/icassp/HuangLTHZIHCW20} & 0.5210 $\pm$ 0.2077 & 0.4851 $\pm$ 0.1849 & 0.4972 $\pm$ 0.1930 & 26.97 & 961.57 & 259.57 & 129.87\\

nnU-Net~\cite{isensee2021nnu} & 0.5047 $\pm$ 0.2002 & 0.4916 $\pm$ 0.1990 & 0.5268 $\pm$ 0.2026 & 18.67 & \textbf{155.01} & \textbf{21.22} & \textbf{10.18}\\

Unlearning~\cite{dinsdale2021deep} & 0.5415 $\pm$ 0.1881 & 0.5632 $\pm$ 0.1721 & 0.5365 $\pm$ 0.1881 & 27.90 & 205.73 & 50.50 & 23.86\\

FAN-Net (ours) & \textbf{0.5591 $\pm$ 0.1801} & \textbf{0.5762 $\pm$ 0.1624} & \textbf{0.5455 $\pm$ 0.1624} & 28.94 & 261.59 & 65.77 & 33.09 \\

\bottomrule

\end{tabular*}
\end{table*}

\begin{figure*}[ht]
\centering
\includegraphics[width=0.95\linewidth]{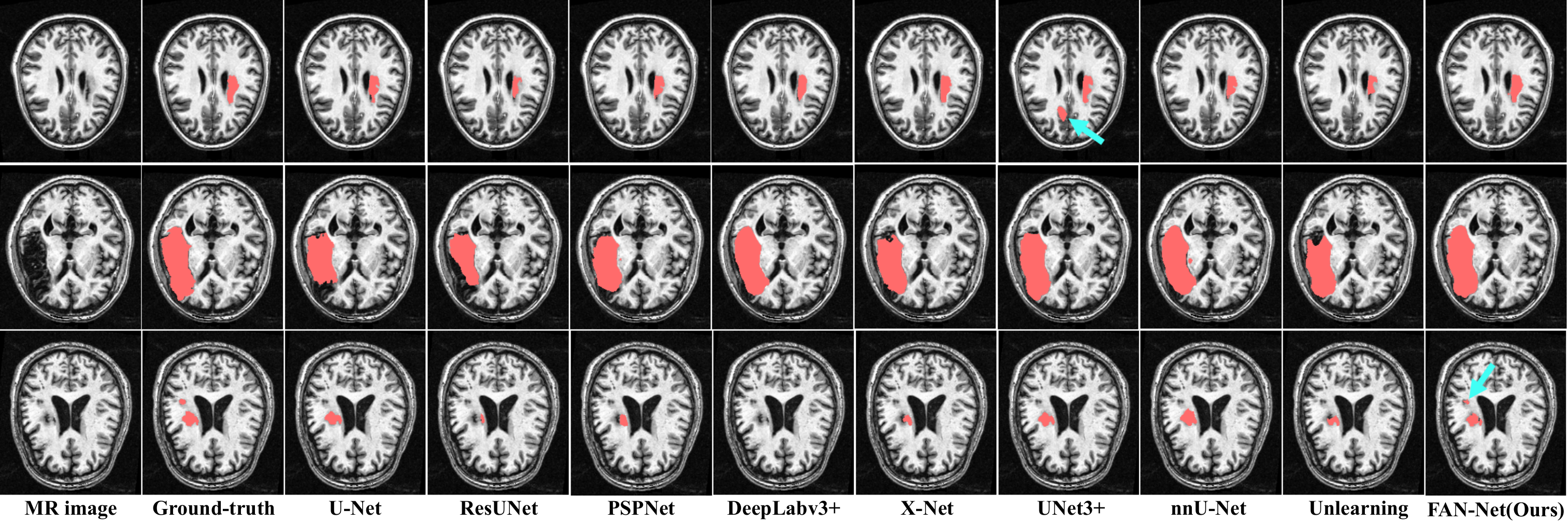}
\vspace{-4mm}
\caption{Examples of segmentation results on ATLAS dataset. Cyan arrows indicate discriminative regions. 
}
\vspace{-4mm}
\label{fig:quality_comparison}
\end{figure*}

Table~\ref{tab:performance_comparison} reports the quantitative results including the mean and standard deviation calculated across 9 independent experiments on 9 split test sets. Obviously, our method has relatively better performance than other methods. 
Moreover, the \methodname can improve the performance without large memory usage or computational costs \textit{w.r.t.} memory, FLOPs, number of parameters, and MACC. Specifically, \methodname outperforms Unlearning\cite{dinsdale2021deep} that is only based on the image space.

For the qualitative comparisons, some segmentation results are shown in Fig.~\ref{fig:quality_comparison}. The third row shows that our method predicts the presence of two small lesions similar to the ground truth, but the other methods can only predict one or none of them. Even in the first row, UNet3+ incorrectly predicts one more lesion.

\subsection{Ablation Studies}\label{sec:exp:ablation}
\noindent\textbf{Effects of $\alpha$ values.}\quad Here, we investigate the effects of the adjusted range of the amplitude component $\alpha$ in Eq.~(\ref{eq:alpha}).
We selected site 5 as the testing set.
According to Table~\ref{tab:alpha}, the suitable setting for the binary mask $M_\alpha$ is $\alpha = 0.1$. If the $\alpha$ value is smaller, the style information to be adjusted is not enough for FAN. While if the value of $\alpha$ is higher, FAN would adjust the high-frequency amplitude component, changing the texture information such as the tissue boundary.

\begin{table}[!t]\small
\renewcommand{\arraystretch}{1.0}
\caption{Performances of using different values of $\alpha$. 
}
\label{tab:alpha}
\centering
\begin{tabular*}{1.0\linewidth}{@{\extracolsep{\fill}}cccc}

\toprule
\bfseries $\alpha$ &\bfseries Dice &\bfseries Recall &\bfseries F1-score\\

\midrule
0.05 & 0.4856 & 0.4948 & 0.5303 \\
0.10 & \textbf{0.5098} & \textbf{0.5117} & \textbf{0.5484} \\
0.15 & 0.4917 & 0.4832 & 0.5406 \\
0.20 & 0.4601 & 0.4586 & 0.4829 \\

\bottomrule
\end{tabular*}
\vspace{-5mm}
\end{table}

\begin{table}[!t]\small
\renewcommand{\arraystretch}{1.0}
\caption{Ablation study on FAN module.
}
\label{tab:ablation}
\centering
\begin{tabular*}{1.0\linewidth}{@{\extracolsep{\fill}}ccccc}

\toprule
\bfseries FAN &\bfseries DC &\bfseries Dice &\bfseries Recall &\bfseries F1-score\\

\midrule
\ & \ & 0.4396 & 0.4044 & 0.4565 \\
\checkmark & \ & 0.4681 & 0.4745 & 0.5082 \\
\checkmark & \checkmark & \textbf{0.5098} & \textbf{0.5117} & \textbf{0.5484} \\

\bottomrule
\end{tabular*}
\vspace{-5mm}
\end{table}

\begin{figure}[t]
\centering
\includegraphics[width=0.85\linewidth]{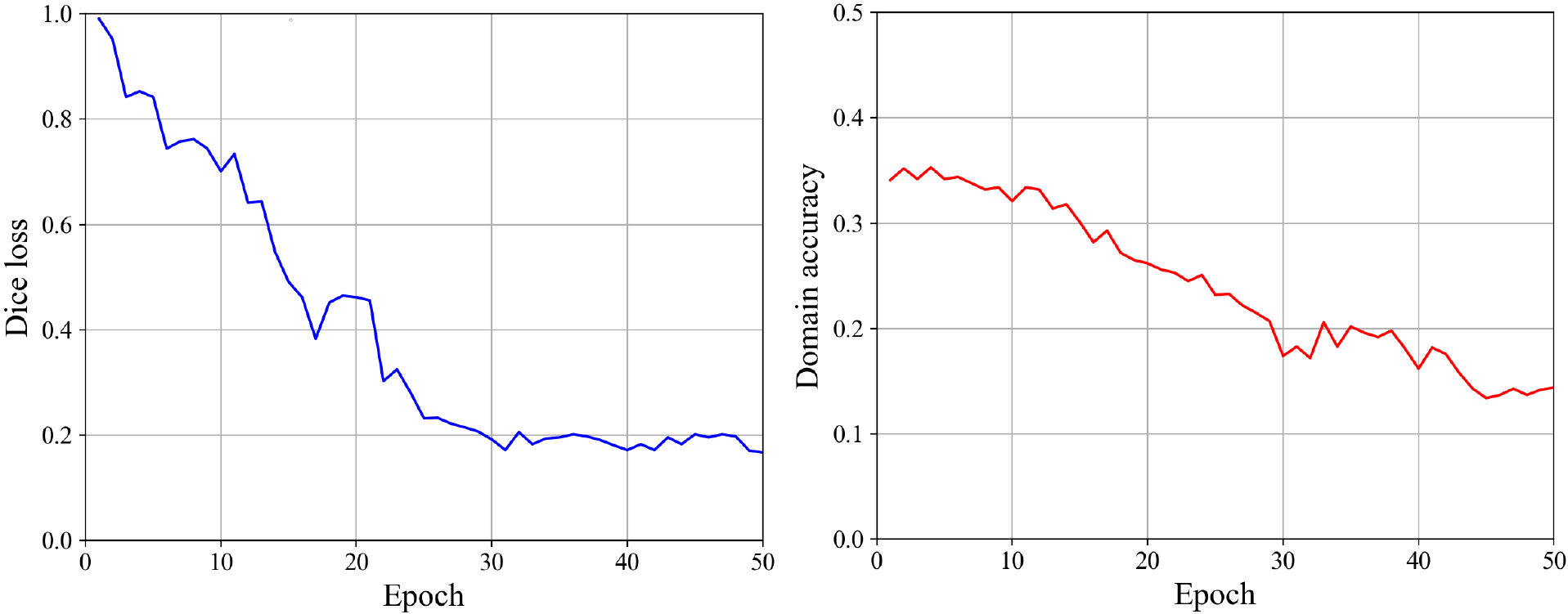}
\vspace{-5mm}
\caption{Curves of the Dice loss (Left) and domain accuracy (Right) along the training.}
\vspace{-4mm}
\label{fig:curve}
\end{figure}

\noindent\textbf{Effects of FAN and DC.}\quad In this experiment, we take U-Net and the z-score--normalized MR images as the baseline, and also select site 5 as the testing set. 
Table~\ref{tab:ablation} shows that each component can consistently improve all metrics. It also demonstrates that the model benefits from the domain-agnostic knowledge obtained by DC. Moreover, in Fig.~\ref{fig:curve} we show the training curves of Dice loss and domain accuracy to understand the behavior of DC. We can see that both of them decrease to a degree, which demonstrates that less domain-specific knowledge is preserved and the model will be more insensitive to the influence of domain diversity and perform stably. Especially, the domain accuracy finally approaches 0.125 (approximately equal to 1/8). On the other hand, this also validates that the learned adaptive affine parameters by FAN definitely model accurate domain knowledge.

\section{Conclusion}\label{sec:conc}

In this paper, we proposed \methodname, which can dynamically change cross-domain style information of MR images without affecting high-level semantic structures. Experimental results on the ATLAS dataset validated the effectiveness of our \methodname, suggesting its potential generalizability to unseen domains in clinical practice.

\clearpage

\end{document}